\documentclass[onecolumn]{jpsj3}
\usepackage{txfonts}
\usepackage{amsmath,amssymb}
\usepackage{graphicx}
\usepackage{endnotes}
\usepackage{color,ulem}
\title{Effects of Interference between Energy Absorption Processes of Molecule and Surface Plasmons on Light Emission Induced by Scanning Tunneling Microscopy}

\author{\name{Kuniyuki \surname{Miwa}}, \name{Mamoru \surname{Sakaue}}, and \name{Hideaki \surname{Kasai}}\thanks{E-mail: kasai@dyn.ap.eng.osaka-u.ac.jp}}
\inst{Department of Applied Physics,
Osaka University, Suita, Osaka 565-0871, Japan}
\abst{
Effects of coupling between an molecular exciton and a surface plasmon (exciton-plasmon coupling) on the luminescence properties of the molecule and the surface plasmons are investigated using nonequilibrium Green's function method. 
Molecular absorption and enhancement by molecular electronic and vibrational modes (molecular modes) lead to dip and peak structures in the luminescence spectra of the surface plasmons. The re-absorption by the surface plasmons leads to a dent structure in their luminescence spectra. It is found that the processes of the molecular absorption and the re-absorption by the surface plasmons interfere with each other. Enhancement and suppression of these energy-absorption processes are due to the interference.
Moreover, due to the exciton-plasmon coupling, excitation channels of the molecule arise in the energy range lower than the first electronic excitation energy of the molecule. It is found that the electron transitions via these excitation channels give rise to the molecular luminescence and the vibrational excitations at the bias voltage lower than the ratio of the first electronic excitation energy of the molecule to the elementary charge. The results also indicate that the vibrational excitations assist the emission of photons, whose energy exceeds the product of the elementary charge and the bias voltage (upconverted luminescence).
}
\kword{molecular luminescence, surface plasmon, scanning tunneling microscopy, molecular vibration, exciton-plasmon coupling, nonequilibrium Green's function method, interference}

\begin{document}
\maketitle

\section{Introduction}
{\color{black}{The optical property of the system, where electronically excited molecules and surface plasmons interact with each other, has been a topic of great interest in various research fields. Surface plasmons localized near metal nanostrcutures generate an intense electromagnetic field in a nano scale region. Therefore, field enhancement by the surface plasmons are commonly employed in many applications, such as surface-enhanced Raman spectroscopy, surface-enhanced infrared absorption spectroscopy, high-resolution biosensors, light-emitting devices, photovoltaics, and so forth.~\cite{Tam2007, Moskovits1985, Nie1997,Pettinger2004, Enders2006, Kundu2008, Lakowicz2001a, Fujiki2010a, Atwater2010, Pillai2010} These pioneering works have mainly focused on how an electromagnetic field generated by surface plasmons affects the optical properties of the target molecules. In addition to this, recent studies suggest that dynamics of the molecules including luminescence and absorption can also affect the optical property of the surface plasmons. For example, coupling of the surface plasmons to a molecular exciton (exciton-plasmon coupling) leads to structures (e.g., peak, dip, and Fano-like structures) in luminescence and absorption spectra of the surface plasmons.~\cite{Ishihara2011,White2012,Schneider2012,Osley2013} Microscopic understanding of effects of the exciton-plasmon coupling will help in elucidating the true nature of the energy transfer between the metal nanostructures and the molecules.\par
%Fundamental understanding of the role of surface plasmons in the molecular luminescence process is currently a topic of great interest owing to its wide applicability in nanoscale spectroscopy, biosensors, light-emitting devices, etc ~\cite{Tam2007,Moskovits1985,Nie1997,Pettinger2004, Lakowicz2001a,Fujiki2010a}.
The highly localized tunneling current of a scanning tunneling microscope (STM) can be utilized as an atomic-scale source to induce light emission from the system. STM-induced light emission (STM-LE) spectroscopy has a unique advantage of having the potential to unveil optical properties with submolecular spatial resolution. Moreover, one of the most intersting things regarding STM-LE from clean and molecule-covered metal substrates is that the surface plasmons localized near the tip-sample gap region contribute to luminescence processes.
\par
%
%STM-LE from a metal surface was first reported by the group of Gimzewski in 1988.~\cite{Gimzewski1988} Now, it is well confirmed that light emission from clean metal surfaces is due to the radiative decay of surface plasmons localized near the tip-sample gap region.~\cite{Berndt1991a,Uehara1999,Hoffmann2001,Rossel2010}
Gimzewski {\textit{et al.}} has first reported STM-LE from a metal surface.~\cite{Gimzewski1988} Nowadays,}} it is well confirmed that light emission from clean metal surfaces is due to the radiative decay of surface plasmons localized near the tip-sample gap region.~\cite{Berndt1991a,Uehara1999,Hoffmann2001, Sakurai2005, Rossel2010}
%%%The plasmon modes depend on the tip-sample geometry. Therefore, changes of the tip shape and/or the tip-sample distance have an influence on the overall shape of the luminescence spectra~\cite{Johansson1990,Johansson1998, Aizpurua2000}.
%\par
When a molecule is {\color{black}{located near}} the gap region, there are two radiative processes, i.e., luminescence from the surface plasmons and the molecule. If the molecule is directly adsorbed on a metal substrate, charge and energy transfer between them lead to the quenching of molecular luminescence. Thus, the molecule acts as a spacer between the tip and the metal substrate, which slightly modifies the plasmon-mediated emission.~\cite{Hoffmann2002,Amemiya2003,Tao2009} If the molecule is electrically decoupled from the metal substrate by dielectric films or molecular multilayers, intrinsic molecular luminescence, which is associated with the electronic and vibrational transitions in the molecule, can be observed.~\cite{Qiu2003,Dong2004,Ino2008,Chen2010}
\par
{\color{black}{Recent studies have reported the importance of }}effects of the interaction between the intense electromagnetic field generated by the surface plasmons and the transition moments for the molecular excitations and de-excitations. Due to the fact that the molecular luminescence can be in the same energy window as the plasmon-mediated emission, light emission can be induced by the two kinds of radiative processes simultaneously. {\color{black}{Uemura {\textit{et al.}} and Liu {\textit{et al.}}}} have shown that the plasmon-mediated emission overlapping with the molecular absorption enhances molecular luminescence intensity.~\cite{Uemura2007,Liu2007a,Liu2009} {\color{black}{Dong {\textit{et al.}} have reported s}}elective enhancement of luminescence intensity by spectrally tuning the surface plasmon mode to match a particular transition in the molecule.~\cite{Dong2010} Moreover, fascinating new phenomena, i.e., luminescence associated with radiative decay from excited vibrational levels for the excited electronic state of the molecule (hot luminescence) and luminescence at energies beyond the product of the elementary charge and the applied bias voltage (upconverted luminescence) have also been observed.~\cite{Dong2010}
\par
{\color{black}{Regarding the excitation processes of the molecule, there are two dominant mechanisms involved.}}
%There are two dominant mechanisms involved in the molecular excitations in STM-LE from molecules on metal substrates.
One is the excitation by the injection of electrons and/or holes from the electrodes. The other is the excitation by the absorption of surface plasmons that are excited by the inelastic tunneling between the tip and the substrate. Recent experiments suggest the importance of the plasmon-mediated excitations ~\cite{Uemura2007,Liu2007a,Liu2009,Dong2010,Zhang2012}. {\color{black}{In surface-plasmon fluorescence spectroscopy, it has reported that the molecule is excited by surface plasmons.~\cite{Ishida1999} Furthermore, a similar plasmon-mediated excitation of molecules has been proposed in previous theoretical studies. In these studies, }}molecular dynamics have been investigated within the framework of the one-body problem, where the surface plasmon is treated as a classical electric field.~\cite{Xu2004,Johansson2005,Tian2011a,Tian2011} Tian \textit{et al.}~\cite{Tian2011a,Tian2011} used the plasmon-mediated excitation mechanism to explain some spectral features observed in Ref.~\citen{Dong2010}. The results indicate that the molecular dynamics induced by the surface plasmons play essential roles.
\par
Direct experimental evidences of the plasmon-mediated excitations of molecules have been obtained in luminescence spectra acquired with tips with a molecule-covered shaft over clean metal surface.~\cite{Schneider2012} Although no electron tunneling to the molecules takes place, the observed spectra, that can be considered as the luminescence spectra of the surface plasmons, show peak and dip structures. {\color{black}{The position of some of these structures matches the position of the peak structures in molecular luminescence and absorption spectra.}} Thus, the molecular dynamics including the molecular luminescence and absorption {\color{black}{can be expected to}} have an impact on luminescence-spectral profiles of the surface plasmons. To understand this from a microscopic point of view, there is a need to investigate the interplay between the dynamics of the molecule and the surface plasmons within the framework of the quantum many-body theory.
\par
We have recently investigated the effects of coupling between a molecular exciton and the surface plasmon (exciton-plasmon coupling) on the luminescence properties with the aid of nonequilibrium Green's function method.~\cite{Miwa2013} The results have first revealed that in addition to the molecular luminescence and absorption, the re-absorption by the surface plasmons modifies the luminescence-spectral profiles of the surface plasmons. It has also been shown that the electron transitions, which give rise to the molecular luminescence and the vibrational excitations, occur at the bias voltage lower than the ratio of the energy of the molecular exciton to the elementary charge $V_\mathrm{bias} < \epsilon_{ex} / e $.~\cite{Miwa2013a}
\par
In this paper, we provide the details on the calculation procedure and the analysis processes. In addition, since the processes of the energy absorption by the molecule and the surface plasmons are found to interfere with each other, the effects of the interference on the luminescence-spectral profiles are discussed. The mechanisms for occurrence of the electronic excitations of the molecule at $V_\mathrm{bias} < \epsilon_{ex} / e $ are also discussed in detail.
\par
%====================================================================
%===== Model ====================
%====================================================================
\section{Model}
We consider the model, where the molecular exciton interacts with a vibron (a molecular-vibrational quantum) and with the surface plasmon.
For simplicity, the molecular vibrations are approximated by a single harmonic oscillator.~\cite{Tian2011} In the present study, three vibrational levels (ground, first-excited and second-excited vibrational levels) are mainly populated. Therefore the harmonic oscillator model will provide valuable insight into the essential physical features.~\cite{foot2,Kasai1993,Kasai2001}
%It should be noted that a more general model for the molecular vibrations, e.g., the one taking into account multiple vibrational modes and/or an intramolecular anharmonicity, can be found in the literature,~\cite{Kasai1993,Kasai2001} and effects of the multiple vibrational modes and/or the intramolecular anharmonicity will be investigated in future works.
The surface plasmon modes localized near the tip-sample gap region are approximated by a single energy mode.~\cite{Tian2011,Tian2011a,Pettinger2007,Berndt1991a} Plasmon mediated emissions acquired with W-tip on clean Au(111) surface and one- or two-monolayers of TPP molecules on Au(111) surface exhibit a single, broad peak in energy range considered in this study.~\cite{Dong2010} Hence higher-order modes, that result in peaks at higher energy range, are ignored.
Creation (annihilation) of the molecular exciton is induced by absorption (emission) of the surface plasmon.~\cite{Tian2011,Tian2011a,Schneider2012,Ishida1999}
Consequently, the Hamiltonian of the system is
\begin{align}
		H  =&  \epsilon_{ex} d^\dagger d
	   		+ \hbar \omega_0 b^\dagger b
	   		+ \hbar \omega_p a^\dagger a
			+ \sum_{\beta} \hbar \omega_\beta b^\dagger_\beta b_\beta
       \nonumber \\
	&   		+ M Q_b d^\dagger d
			+ V \left( a d^\dagger + d a^\dagger \right)
	   		+ \sum_{\beta} U_\beta Q_b Q_\beta ,
	\label{Hami}
\end{align}
\noindent where $d^\dagger$ and $d$ are creation and annihilation operators (for later convenience assumed to satisfy boson commutation rules) for the molecular exciton with energy $\epsilon_{ex}$. Operators $b^\dagger$ and $b$ are creation and annihilation operators for the molecular vibrational mode with energy $ \hbar \omega_0 $, $a^\dagger$ and $a$ are for a surface plasmon mode with energy $ \hbar \omega_p $, $b^\dagger_\beta$ and $b_\beta$ are for a phonon mode in a thermal phonon bath, $Q_b=b+b^\dagger$ and $Q_\beta=b_\beta+b^\dagger_\beta$.
The energy parameters $M$, $V$ and $U_\beta$ correspond to the dipolar exciton-vibron coupling, the exciton-plasmon coupling, and the coupling between the vibron and the phonon in the thermal phonon bath.
The model system Eq.~(\ref{Hami}) is a generalized one that can be utilized to investigate STM-LE from molecular layers~\cite{Dong2010} and from a clean metal surface with a molecule-covered tip.~\cite{Schneider2012} Since the parameter values are expected to be of the same order in the situation where the molecules are on molecular layers or adsorbed on a tip surface, the same model system Eq.~(\ref{Hami}) is applied in both cases. 
\par
By applying the Lang-Firsov (LF) canonical transformation,~\cite{Galperin2006b}
$\Tilde{H}=e^S H e^{-S}$ with $S=\lambda \left( b^\dagger - b \right) d^\dagger d$,
$H$ becomes
\begin{align}
	\Tilde{H}
		=& \epsilon_{ex} d^\dagger d
	   		+ \hbar \omega_0 b^\dagger b
	   		+ \hbar \omega_p a^\dagger a
	   		+ \sum_{\beta} \hbar \omega_\beta b^\dagger_\beta b_\beta
	   		\nonumber \\
	   	 & + V \left(  a  d^\dagger       X^\dagger
	   		          + \mathrm{H.c.} \right)
	   		+ \sum_{\beta} U_\beta Q_b Q_\beta ,
	   		\label{LFtransform}
\end{align}
where
$ X  = \exp \left[ - \lambda \left( b^\dagger - b \right) \right]$ and $\lambda = M / (\hbar \omega_0)$. Here, the term including $d^\dagger d d^\dagger d$ is neglected. This approximation is expected to be valid, since the population of the molecular exciton is low $\left( <10^{-4} \right)$.\par
The luminescence spectra of the molecule and the surface plasmons are obtained by the relations,~\cite{Galperin2009}
\begin{align}
	J_L(\omega) =& -\frac{1}{\pi} \Im L^< (\omega) ,	\\
	J_P(\omega) =& -\frac{1}{\pi} \Im P^< (\omega) .
\end{align}
Here $L^<$ and $P^<$ are the lesser projections of the Green's function of the molecular exciton and the surface plasmon on the Keldysh contour ($L$ and $P$), respectively.
In STM-LE from molecular layers and from a clean metal surface with a molecule-covered tip, quenching of molecular luminescence is suppressed and relatively significant, respectively. When the quenching is suppressed, $J_L$ is expected to be detected experimentally.~\cite{Dong2004}
When the quenching is relatively significant, the luminescence spectra detected experimentally are expected to correspond to $J_P$.~\cite{foot}
\par
To investigate the excitation properties and analyze the structures in the luminescence spectra, the spectral functions of the molecule and the surface plasmons, which correspond to their absorption spectra, are calculated by the relations,
\begin{align}
	\rho_L(\omega) =& -\frac{1}{\pi} \Im L^r (\omega) ,	\\
	\rho_P(\omega) =& -\frac{1}{\pi} \Im P^r (\omega) ,
\end{align}
where $L^r$ and $P^r$ are the retarded projections of $L$ and $P$.
\par
To obtain the spectral functions and the luminescence spectra, it is needed to calculate the LF transformed $L$ and $P$, which are defined as
\begin{align}
L(\tau,\tau')
	&=  \frac{1}{i \hbar} \langle T_C \{ d (\tau) X(\tau) d^\dagger (\tau') X^\dagger(\tau') \} \rangle _{\Tilde{H}} ,
	\label{MoleGre} \\
P(\tau,\tau')
	&=  \frac{1}{i \hbar} \langle T_C \{ a(\tau) a^\dagger(\tau') \} \rangle _{\Tilde{H}} ,
\end{align}
where $\langle\cdots\rangle _{\Tilde{H}}$ denotes statistical average in representation by system evolution for $\Tilde{H}$. The variable $\tau$ is the Keldysh contour time variable. The operator $T_C$ denotes the time ordering along the Keldysh contour.~\cite{Keldysh1965}
The explicit appearance of the operators $X$ and $X^\dagger$ in the LF transformed Green's function of the molecular exciton (polarization propagator) Eq.~(\ref{MoleGre}) arise from the dipolar exciton coupling to the vibron. This is in contrast to the case of equivalent monopole electron and hole coupling to local boson studied in Ref.~\citen{Gumhalter1984}, Sec.~5C, in which these terms are absent from the polarization propagator.
\par
\begin{figure}
\centering
\includegraphics[ width = 8.6cm, angle = 0] {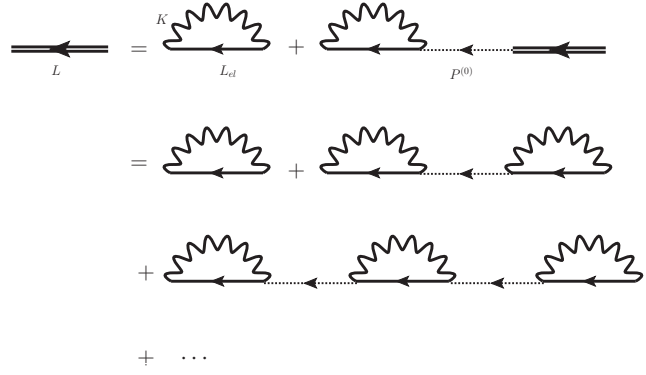}%
\caption{\label{Diagram_RPA} Diagrammatic representation of Eq.~(\ref{eq:polarization}). Heavy double, heavy solid wiggly and dotted lines are $L$, $L_{el}$, $K$ and $P^{(0)}$, respectively.}
\end{figure}
\begin{figure}
\centering
\includegraphics[ width = 8.6cm, angle = 0] {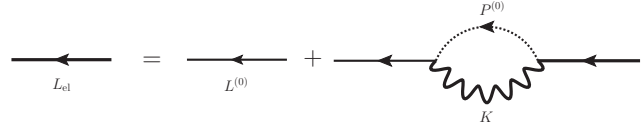}%
\caption{\label{Diagram_Decouple} Diagrammatic representation of Eq.~(\ref{eq:L_ele}). Heavy solid, wiggly, dotted and solid lines are $L_{el}$, $K$, $P^{(0)}$ and $L^{(0)}$, respectively.}
\end{figure}

To calculate the Green's function $L$ containing the exciton-plasmon coupling $V$, we take into account a random-phase-approximation (RPA)-type diagrammatic series. This approximation is expected to hold for $M \gg V^2$.~\cite{Maier2011} The integral equation for $L$ is given by
\begin{align}
	L(\tau,\tau')
	 =& L_\mathrm{b} (\tau,\tau') \nonumber \\
	   & +\int d\tau_1 d\tau_2 L_\mathrm{b} (\tau,\tau_1) |V|^2 P^{(0)}(\tau_1, \tau_2) L(\tau_2,\tau'),
	\label{eq:polarization}~~
\end{align}
where
\begin{align}
	L_\mathrm{b} (\tau,\tau') &= L_\mathrm{el}(\tau,\tau') K(\tau,\tau'),
	\label{eq:decoupling} \\
	L_\mathrm{el} (\tau,\tau') &= \frac{1}{i\hbar} \langle T_C \{ d (\tau) d^\dagger (\tau') \}\rangle_{\Tilde{H}}, \\
	K(\tau,\tau') &= \langle T_C \{ X(\tau) X^\dagger(\tau') \} \rangle_{\Tilde{H}},
\end{align}
{\noindent}and $P^{(0)}$ is the plasmon Green's function for $V=0~\mathrm{eV}$.  The diagrammatic representation of Eq.~(\ref{eq:polarization}) is given in Fig.~\ref{Diagram_RPA}.
%Figure%%%%%
%%%%%%%%%%
Here, we employ an approximation of decoupling electronic and vibrational parts in $L_\mathrm{b}$, and calculate $L_\mathrm{el}$ and $K$. The integral equation for $L_\mathrm{el}$ is given by
\begin{align}
	L_\mathrm{el}(\tau,\tau')
	=& L^{(0)}(\tau,\tau') \nonumber \\
	&+ \int d\tau_1 d\tau_2
		  L^{(0)}(\tau,\tau_1)
		  \Sigma (\tau_1, \tau_2)
		  L_\mathrm{el}(\tau_2,\tau'),~ ~
	\label{eq:L_ele}\\
	\Sigma (\tau_1, \tau_2)
	=& |V|^2 P^{(0)} (\tau_1, \tau_2) K(\tau_2,\tau_1),
\end{align}
where $L^{(0)}$ is $L_\mathrm{el}$ for $V=0~\mathrm{eV}$. The diagrammatic representation of Eq.~(\ref{eq:L_ele}) is shown in Fig.~\ref{Diagram_Decouple}.
%Figure%%%%%
%%%%%%%%%%
\par
%===== end Exciton Green's function =====
%===== begin Plasmon Green's function =====
The integral equation for the plasmon Green's function $P$ is
\begin{align}
	P(\tau,\tau')
	=& P^{(0)}(\tau,\tau') \nonumber \\
	&+ \int d\tau_1 d\tau_2
		  P^{(0)}(\tau,\tau_1) |V|^2 L (\tau_1, \tau_2) P(\tau_2,\tau').~~~~
	\label{eq:Pla}
\end{align}
\par
%===== end Exciton Green's function =====
%===== begin Vibron Green's function =====
As for the correlation function $K$, it is expressed in terms of the vibrational momentum Green's function $D$, which is defined by
\begin{equation}
	D(\tau,\tau') = \frac{1}{i\hbar}
	    \langle T_C\{ P_b (\tau) P_b (\tau') \}\rangle_{\Tilde{H}}
\end{equation}
where $P_b=-i\left(b-b^\dagger\right) $. For the second order cumulant expansion of $\lambda$, this connection is given by
\begin{equation}
       K(\tau,\tau\rq{}) = \exp\left\{ i \hbar \lambda^2 \left[ D(\tau,\tau') - D(\tau,\tau) \right] \right\}.
       \label{eq:shift}
\end{equation}
The integral equation for $D$ is given by
\begin{align}
       D(\tau,\tau') =& D^{(0)}(\tau,\tau') \nonumber \\
	                   &+ \int d\tau_1 d\tau_2 D^{(0)}(\tau,\tau_1)\Pi_\mathrm{el} (\tau_1, \tau_2) D(\tau_2,\tau'),~~~~
		  \label{eq:D}\\
	\Pi_\mathrm{el}  (\tau, \tau')
	=& i \hbar \lambda^2 |V|^2 P^{(0)} (\tau, \tau') K(\tau',\tau)
	\nonumber \\
	& + i \hbar \lambda^2 |V|^2 K(\tau, \tau') P^{(0)} (\tau',\tau) .
	\label{eq:Pi}
\end{align}
Here, $D^{(0)}$ is the vibrational momentum Green's function for $V=0~\mathrm{eV}$. It contains the self-energy emerged from the coupling to the thermal phonon bath $U_\beta$.~\cite{Galperin2006b}
\par
%===== end Vibron Green's function =====
%===== begin how to calculate Green's function =====
By assuming the condition of stationary current, the distribution function $N_\mathrm{pl}$ of the surface plasmons excited by inelastic tunneling between the tip and the substrate is given by
\begin{equation}
	N_\mathrm{pl}(\omega)
	= \begin{cases}
	T_\mathrm{pl} \left(1-\left| \frac{ \hbar \omega }
	                     { eV_\mathrm{bias} } \right| \right),
	& \left| \hbar \omega \right| <
	              \left| eV_\mathrm{bias} \right| \\
    0,	& \text{others}
	  \end{cases}
	,
	\label{N_Pla}
\end{equation}
where $T_\mathrm{pl}$ is a coefficient related to the current amplitude due to the inelastic tunneling,~\cite{Persson1992} $e$ is the elementary charge, and $V_\mathrm{bias}$ is the bias voltage.
The lesser and greater projections of $P^{(0)}$ are obtained through the relations, $P^{(0),<} (\omega)=2iN_\mathrm{pl}(\omega)\Im{P^{(0),r}(\omega)}$ and $P^{(0),>} (\omega)=2i\left[ 1 + N_\mathrm{pl}(\omega) \right] \Im{P^{(0),r}}$, where $P^{(0),r}$ is the retarded projection of $P^{(0)}$.~\cite{Keldysh1965}
The equations mentioned above are self-consistently solved as described in the next section.
\par
%====================================================================
%===== Calculation Scheme =========
%====================================================================
\section{Self-Consistent Calculation Scheme}
Due to the connection between the correlation function $K$ and the vibrational momentum Green's function $D$, represented by Eq. (\ref{eq:shift}), it is easier to express the lesser and greater projections of $K$ in terms of $D$ in the time domain. At the same time, expressions for the lesser and greater projections of $L$, $L_\mathrm{el}$, $D$, and $P$ are easier to be written in the energy domain. Hence, we are working in both time and energy domains, by implementing a fast Fourier transform between them. The ensuring self-consistent calculation scheme consists of the following steps:
\par
1. We start with the retarded projection of the Green\rq{}s functions in the energy domain for $V=0~\mathrm{eV}$
\begin{align}
	D^{(0),r}(\omega)
	=& \left[ \hbar \omega - \hbar \omega_0 +i \frac{\hbar \gamma}{2} \right]^{-1}
	- \left[ \hbar \omega + \hbar \omega_0 +i \frac{\hbar \gamma}{2} \right]^{-1},
	\nonumber \\
	\\
%\end{align}
%\begin{equation}
	P^{(0),r}(\omega)
	=& \left[ \hbar \omega - \hbar \omega_p +i \frac{\hbar \Gamma_\mathrm{pl}}{2} \right]^{-1},
%\end{equation}
\end{align}
where $\hbar \gamma$ represents the vibrational damping due to $U_\beta$, which is given by
\begin{equation}
	\hbar \gamma
	= 2\pi \sum_\beta |U_\beta|^2 \delta \left( \hbar \omega - \hbar \omega_\beta \right),
\end{equation}
and $\hbar \Gamma_\mathrm{pl}$ represents the decay of the surface plasmons for $V=0~\mathrm{eV}$.
The lesser and greater projections of these Green's functions are obtained by
\begin{align}
	D^{(0),<}(\omega)
	=& 2i F(\omega) \Im {D^{(0),r}(\omega)} ,\\
	D^{(0),>}(\omega)
	=& 2i F(-\omega) \Im {D^{(0),r}(\omega)}, \\
	P^{(0),<}(\omega)
	=& 2 i N_\mathrm{pl} (\omega) \Im{P^{(0),r}(\omega)},
	\\
	P^{(0),>}(\omega)
	=& 2 i \left[ 1+N_\mathrm{pl} (\omega) \right] \Im{P^{(0),r}(\omega)},
\end{align}
where
\begin{align}
       F(\omega)
	= \begin{cases}
	N_\mathrm{BE}(|\omega|),
	& \omega > 0\\
       1+N_\mathrm{BE}(|\omega|),
	& \omega < 0
	  \end{cases}
	.
	\label{N_BE}
\end{align}
Here, $N_\mathrm{BE}(\omega)$ is the Bose-Einstein distribution function. The lesser and greater projections are then transformed to the time domain by the Fourier transformation;
\begin{align}
       G^{<,>}(t)
	= \int_{-\infty}^{\infty} \frac{d\omega}{2\pi}G^{<,>}(\omega)e^{-i\omega t},
	\label{Fourier}
\end{align}
where $G^{<,>}$ represent the lesser and greater projections of the Green\rq{}s functions $G$, and $t$ is the real time variable.
\par
2. By utilizing the Langreth theorem~\cite{Haug2007} and using $D$ in lesser and greater projections of Eq.(\ref{eq:shift}), the lesser and greater correlation functions for $K$ can be obtained with~\cite{Galperin2006b}
\begin{align}
       K^{<}(t)
	 = & \langle X^\dagger(0) X(t) \rangle_{\Tilde{H}} \nonumber \\
	 = & \exp\left\{ i \hbar \lambda^2 \left[ D^<(t) - D^<(t=0) \right] \right\}, \\
	K^{>}(t)
	= & \langle X(t) X^\dagger(0) \rangle_{\Tilde{H}} \nonumber \\
	= & \exp\left\{ i \hbar \lambda^2 \left[ D^<(t) - D^<(t=0) \right] \right\}.
\end{align}
In the first iteration step, $D=D^{(0)}$ is applied.
\par
% set_new_ExGF
3. The obtained $K^{<,>}(t)$ are convoluted with $P^{(0)}$ to obtain $\Sigma$ according to the relations
\begin{align}
       \Sigma^<(t)
	 = & |V|^2 P^{(0),<}(t) K^>(-t), \\
	\Sigma^>(t)
	 = & |V|^2 P^{(0),>}(t) K^<(-t).
\end{align}
The lesser and greater projections of $\Sigma$ are then transformed to the energy domain by the inverse Fourier transformation. They are used for calculating the retarded projection of $\Sigma$.
\par
4. The self-energies $\Sigma$, obtained in the previous step, are used for calculating $L_\mathrm{el}$ (retarded, lesser and greater projections) in the energy domain. The retarded projection of $L_\mathrm{el}$ are
\begin{align}
       L^r_\mathrm{el} (\omega) = \left[ \hbar \omega - \epsilon_{ex} -\Sigma^r(\omega) \right]^{-1}.
\end{align}
Similarly, by utilizing the Langreth theorem and the inverse Fourier transformation, the lesser and greater projections of $L_\mathrm{el}$ are obtained in the energy domain.
%\begin{align}
 %      L^<_\mathrm{el}(\omega) &=& L^r_\mathrm{el}(\omega) \Sigma^<(\omega) L^a_\mathrm{el}(\omega),
%       \\
%       L^>_\mathrm{el}(\omega) &=& L^r_\mathrm{el}(\omega) \Sigma^>(\omega) L^a_\mathrm{el}(\omega).
%\end{align}
Using $K$ and $L_\mathrm{el}$ in lesser and greater projections of Eq.(\ref{eq:decoupling}) and the Langreth theorem, the lesser and greater projections of $L_\mathrm{b}$ can be obtained as
\begin{align}
       L^<_\mathrm{b}(t)
	 = &  L^<_\mathrm{el} (t) K^<(t), \\
	L^>_\mathrm{b}(t)
	 = &  L^>_\mathrm{el} (t) K^>(t).
\end{align}
These functions are then transformed to the energy domain and used for calculating the retarded projection of $L_\mathrm{b}$.
\par
% set_new_LeGF
5. The obtained $L_\mathrm{b}$ and $P^{(0)}$ are used for calculating $L^r$, according to Eq.(\ref{eq:polarization}). The retarded projection of $L$ is
\begin{align}
       L^r (\omega) = \left\{ \left[ L^r_\mathrm{b}(\omega) \right]^{-1} -|V|^2 P^{(0),r}(\omega) \right\}^{-1}.
\end{align}
Utilizing the Langreth theorem and the inverse Fourier transformation, the lesser and greater projections of $L$ become
\begin{align}
       L^<(\omega)
       =& L^r(\omega) \frac{L^<_\mathrm{b}(\omega)}{L^r_\mathrm{b}(\omega)}
       \nonumber \\
       &    
               + L^r(\omega) |V|^2 P^{(0),<}(\omega) L^a(\omega)
       \nonumber \\
       &
               + L^r(\omega) |V|^2 \frac{L^<_\mathrm{b}(\omega)}{L^r_\mathrm{b}(\omega)} P^{(0),a}(\omega) L^a(\omega) , \\
       L^>(\omega)
       =& L^r(\omega) \frac{L^>_\mathrm{b}(\omega)}{L^r_\mathrm{b}(\omega)}
       \nonumber \\
       &
              + L^r(\omega) |V|^2 P^{(0),>}(\omega) L^a(\omega)
       \nonumber \\
       &
              + L^r(\omega) |V|^2 \frac{L^>_\mathrm{b}(\omega)}{L^r_\mathrm{b}(\omega)} P^{(0),a}(\omega) L^a(\omega) .
\end{align}
% set_new_PlaGF
The obtained $L^{r,<,>}$ are used for calculating $P$, according to Eq. (\ref{eq:Pla}).
The retarded, lesser and greater projections of $P$ are
\begin{align}
       P^r (\omega) = \left\{ \left[ P^{(0),r}(\omega) \right]^{-1} -|V|^2 L^r(\omega) \right\}^{-1},
\end{align}
\begin{align}
       P^<(\omega)
       =& P^r(\omega) \frac{P^{(0),<}(\omega)}{P^{(0),r}(\omega)}
       \nonumber \\
       &
              + P^r(\omega) |V|^2 L^<(\omega) P^a(\omega)
       \nonumber \\
       &
              + P^r(\omega) |V|^2 \frac{P^{(0),<}(\omega)}{P^{(0),r}(\omega)} L^a(\omega) P^a(\omega),
       \\
       P^>(\omega)
       =& P^r(\omega) \frac{P^{(0),>}(\omega)}{P^{(0),r}(\omega)}
       \nonumber \\
       &
              + P^r(\omega) |V|^2 L^>(\omega) P^a(\omega)
       \nonumber \\
       &
              + P^r(\omega) |V|^2 \frac{P^{(0),>}(\omega)}{P^{(0),r}(\omega)} L^a(\omega) P^a(\omega).
\end{align}
In this step, $L$ and $P$ are transformed to the time domain. They are substituted into the lesser and greater projections of Eq.(\ref{eq:Pi}) to obtain $\Pi^{<,>}_{el}(t)$. They are then transformed to the energy domain and used for calculating $\Pi^{r}_{el}(\omega)$.
\par
6. Then $D$ is updated by Eq. (\ref{eq:D}). The retarded, lesser and greater projections of $D$ are
\begin{align}
       D^r (\omega) = \left\{ \left[ D^{(0),r}(\omega) \right]^{-1} -\Pi^{r}_{el}(\omega) \right\}^{-1},
\end{align}
\begin{align}
       D^<(\omega)
       =& D^r(\omega) \frac{D^{(0),<}(\omega)}{D^{(0),r}(\omega)}
       \nonumber \\
       &
        + D^r(\omega) \Pi^<_{el}(\omega) D^a(\omega)
       \nonumber \\
       &
        + D^r(\omega) \frac{D^{(0),<}(\omega)}{D^{(0),r}(\omega)} \Pi^a_{el}(\omega)  D^a(\omega), \\
       D^>(\omega)
       =& D^r(\omega) \frac{D^{(0),>}(\omega)}{D^{(0),r}(\omega)}
       \nonumber \\
       &
              + D^r(\omega) \Pi^>_{el}(\omega) D^a(\omega)
       \nonumber \\
       &
              + D^r(\omega) \frac{D^{(0),>}(\omega)}{D^{(0),r}(\omega)} \Pi^a_{el}(\omega) D^a(\omega).
\end{align}
They are used again in step 2. Hence, $K$, $\Sigma$, $L_\mathrm{el}$, $L_\mathrm{b}$, $L$, $P$ and $\Pi_{el}$ are updated in the same way. The above procedure is repeated until the proportion of the absolute change of the population of the molecular exciton $n_e$ in two subsequent iterations to the latest $n_e$ is less than a predefined threshold of $10^{-7}$. Here, $n_e$ is given by
\begin{equation}
	n_e = - \Im \left[ L^< (t=0) \right].
\end{equation}
\par
After convergence, the spectral functions and the luminescence spectra are calculated. To analyze the luminescence spectra, the lesser projections of the right in Eqs. (\ref{eq:polarization}) and ({\ref{eq:Pla}) are divided into four terms using the Langreth theorem.
%The lesser projections of Eqs. (\ref{eq:polarization}) and ({\ref{eq:Pla}) in the energy domain are given by
\begin{align}
L^<(\omega)
	=& L^<_\mathrm{b} (\omega) \nonumber \\
	 & +|V|^2 L^<_\mathrm{b} (\omega) P^{(0),a}(\omega) L^a(\omega) \nonumber \\
	 & +|V|^2 L^r_\mathrm{b} (\omega) P^{(0),<}(\omega) L^a(\omega) \nonumber \\
	 & +|V|^2 L^r_\mathrm{b} (\omega) P^{(0),r}(\omega) L^<(\omega),
   \label{eq:EachML}
\end{align}
\begin{align}
P^<(\omega)
	=& P^{(0),<} (\omega) \nonumber \\
	 & +|V|^2 P^{(0),<} (\omega) L^a (\omega) P^a(\omega) \nonumber \\
	 & +|V|^2 P^{(0),r} (\omega)  L^<(\omega) P^a(\omega) \nonumber \\
	 & +|V|^2 P^{(0),r} (\omega)  L^r (\omega) P^<(\omega).
   \label{eq:EachPL}
\end{align}
\par
The parameters given correspond to the experiment on the STM-LE from tetraphenylporphyrin(TPP) molecules on gold surface:~\cite{Dong2010} $\epsilon_{ex} = 1.89~\mathrm{eV}$, $\hbar \omega_0 = 0.16~\mathrm{eV}$ and $\hbar\omega_p=2.05~\mathrm{eV}$. The statistical average is taken for temperature $T=80~\mathrm{K}$.~\cite{Dong2010} The square of $\lambda$ is reported to be 0.61 on the basis of first-principles calculations.~\cite{Tian2011} The parameter $U_\beta$ is given so that the molecular vibrational lifetime due to the coupling to the thermal phonon bath is 13 ps.~\cite{Dong2010} A Markovian decay is assumed for the surface plasmon to give a plasmon lifetime for $V=0~\mathrm{eV}$ of 4.7 fs.~\cite{Dong2010,Tian2011} The coefficient $T_\mathrm{pl}$ is set to $10^{-4}$, where a tunneling current is $I_t=200~\mathrm{pA}$ and an excitation probability of the surface plasmons per electron tunneling event is considered to be $10^{-2}$.
%====================================================================
%===== Results and Discussion =========
%====================================================================
\section{Results and Discussion}
%====================================================================
%===== Plasmon =========
%====================================================================
\subsection{Plasmon luminescence}
Figure~\ref{PL_Bias} shows the luminescence spectra of the surface plasmons $J_P$ for various bias voltages $V_\mathrm{bias}$.
%Figure ( Bias voltage dependence)%%%%%%%%%%%%%%%%%%%%%%%%%
\begin{figure}
\centering
\includegraphics[ width = 8.6cm] {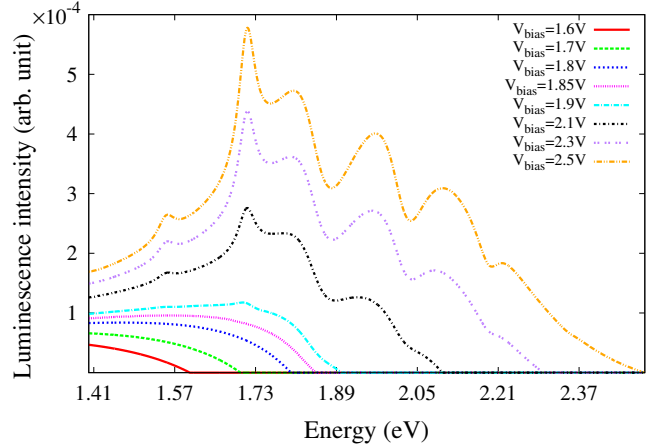}%
\caption{\label{PL_Bias} (Color online) Bias voltage $V_\mathrm{bias}$ dependence of the luminescence spectra of the surface plasmon $J_P$. Red solid, green dashed, blue dotted, magenta dotted, cyan dashed-dotted, black dashed-dotted, purple dotted and orange dashed double-dotted lines are $J_P$ at $V_\mathrm{bias}$=1.6~V, 1.7~V, 1.8~V, 1.85~eV, 1.9~V, 2.1~V, 2.3~V and 2.5~V.}
\end{figure}
%%%%%%%%%%%%%%%%%%%%%%%%%%%%%%%%%%%%%%%%
At $V_\mathrm{bias} \le 1.85~\mathrm{V}$, $J_P$ shows a smooth curve. At $V_\mathrm{bias} > 1.85~\mathrm{V}$, complicated peak and dip structures appear.
%Figure ( Bias voltage dependence)%%%%%%%%%%%%%%%%%%%%%%%%%
\begin{figure}
\begin{center}
\includegraphics[ width = 7cm] {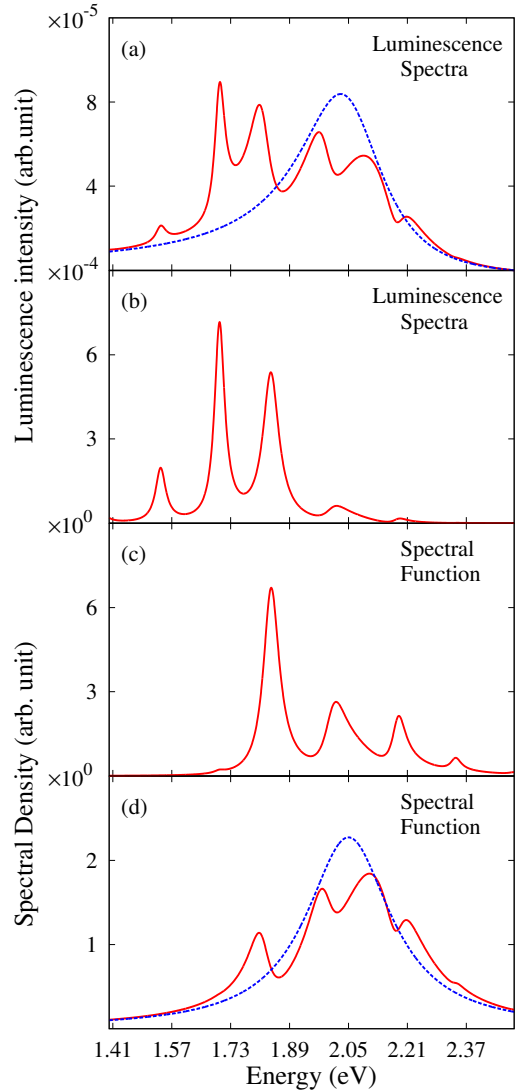}%
\caption{\label{PL_Comp} (Color online) (a) Luminescence spectra of the surface plasmons $J_P$, (b) luminescence spectra of the molecule $J_L$, (c) spectral function of the molecule $\rho_L$ and (d) spectral function of the surface plasmons $\rho_P$. Red solid and blue dashed lines represent the result for $V=0.10~\mathrm{eV}$ and $V=0.00~\mathrm{eV}$, respectively. The bias voltage is $V_\mathrm{bias}=2.5~\mathrm{V}$}
\end{center}
\end{figure}
%%%%%%%%%%%%%%%%%%%%%%%%%%%%%%%%%%%%%%%%
The position of the peak structures in the energy range $\hbar \omega<1.75~\mathrm{eV}$ (near 1.71~eV and 1.55~eV) correspond to that of the peak structures in the luminescence spectra of the molecule $J_L$ [Fig.~\ref{PL_Comp}(b)]. The position of the dip structures in the energy range $\hbar \omega > 1.8~\mathrm{eV}$ (near 1.88~eV, 2.03~eV and 2.19~eV) correspond to that of the peak structures in the spectral function of the molecule $\rho_L$ [Fig.~\ref{PL_Comp}(c)]. A dent structure appears near the energy of the surface plasmon mode (2.05~eV), where the spectral function of the surface plasmons $\rho_P$ has its maximum intensity [Fig.~\ref{PL_Comp}(d)].
\par
To analyze the origin of these structures in $J_P$, contributions from each term in Eq. (\ref{eq:EachPL}) are shown in Fig.~\ref{PL_Each}.
%Figure (Contribution from each term) %%%%%%%%%%%%%%%%%%%%
\begin{figure}
\centering
\includegraphics[ width = 8.6cm] {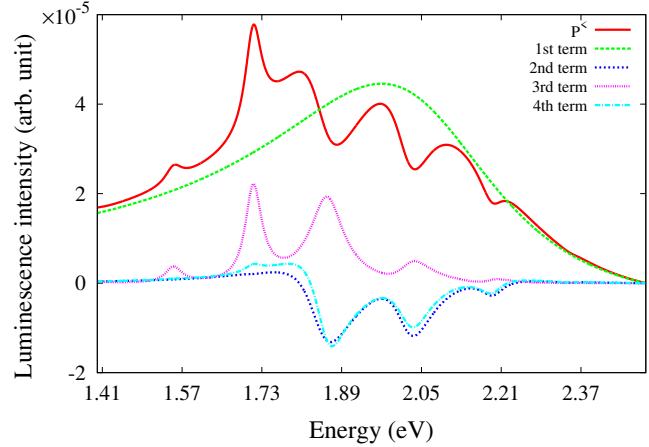}%
\caption{\label{PL_Each} (Color online) Imaginary part of each term in Eq. (\ref{eq:EachPL}). Red solid line is the luminescence spectra of the surface plasmons $J_P$. Green dashed, blue dotted, magenta dotted and cyan dashed-dotted lines are the product of $-1/\pi$ and the imaginary part of the first, second, third and fourth terms on the right side of Eq. (\ref{eq:EachPL}), respectively. The bias voltage is $V_\mathrm{bias}=2.5~\mathrm{V}$}
\end{figure}
%%%%%%%%%%%%%%%%%%%%%%%%%%%%%%%%%%%%%%%%
The contribution from the first term on the right side of Eq. (\ref{eq:EachPL}) (green dashed line) represents the plasmon luminescence where the exciton-plasmon coupling is ignored, i.e., $V=0~\mathrm{eV}$. The third term (magenta dotted line) expresses the contribution of the energy transfer from the molecular exciton. The luminescence spectra of the surface plasmon $J_P$ (red line) shows peak structures near 1.71~eV and 1.55~eV. These peak structures are, therefore, attributed to enhancement by the molecular electronic and vibrational modes (molecular modes). The second and fourth terms (magenta dotted and cyan dashed-dotted lines) give the processes where the energy of the surface plasmons are absorbed by the molecule and the surface plasmons (molecular absorption and re-absorption by the surface plasmons, respectively). For the contribution from the second and fourth terms, the dip structures appear near 1.87~eV, 2.03~eV and 2.19~eV in $J_L$. Since the spectral intensity in the spectral function of the surface plasmons $\rho_P$ is high in the energy range from 1.85~eV to 2.2~eV, the re-absorption by the surface plasmons leads to the dent structure in this energy range.
\par
{\color{black}{However, the origin of the structure near 1.81~eV in $J_P$ (red line in Fig.~\ref{PL_Each}), which looks like a peak structure, cannot be explained by each of these processes. There is a need to consider a superposition of the processes of the molecular absorption and the re-absorption by the surface plasmons. To analyze this,}} the real and imaginary parts of $P^{(0),r}$ and $L^r$ are shown in Fig.~\ref{Spectra}.
%Figure (Real and imaginary parts of L and P) %%%%%%%%%%%%%%%%%%%%
\begin{figure}
\centering
\includegraphics[width=8.5truecm]{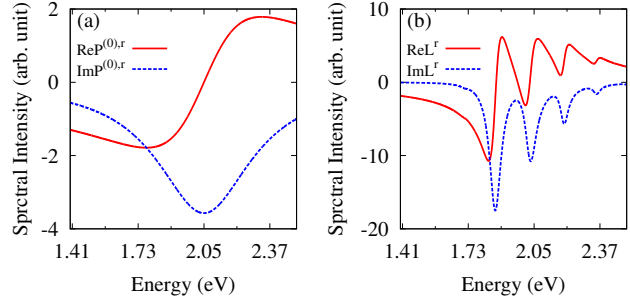}
\caption{\label{Spectra} (Color online) Calculation results for (a) the real and imaginary parts of $P^{(0),r}$ and (b) the real and imaginary parts of $L^r$. (a) Red solid and blue dashed lines are the real and imaginary parts of $P^{(0),r}$, respectively. (b) Red solid and blue dashed lines are the real and imaginary parts of $L^r$, respectively.
}
 \end{figure}
%%%%%%%%%%%%%%%%%%%%%%%%%%%%%%%%%%%%%%%%%%%%%%%%%%%%%%%%%%%%%
The product of $P^<$ and $\left( \Re{P^{(0),r}} \Re{L^r} - \Im{P^{(0),r}} \Im{L^r}\right)$ gives the fourth term on the right side of Eq. (\ref{eq:EachPL}). As is well known, the lesser projection $P^<$ is a pure imaginary number in the energy domain.~\cite{Keldysh1965} The function $-\Im{P^{(0),r}}/\pi$ gives the spectral function of the surface plasmons for $V=0~\mathrm{eV}$, and $-\Im{L^r}/\pi$ gives the spectral function of the molecule. These functions correspond to their absorption spectra.\par
{\color{black}{One of the possible mechanisms for the occurrence of the structure near 1.81~eV in $J_P$ is an interference between the processes of the molecular absorption and the re-absorption by the surface plasmons. To investigate the interference effects, the real parts of $P^{(0),r}$ and $L^r$, the product of which leads to the interference between these processes, are analyzed. Since the real and imaginary parts of $P^{(0),r}$ satisfy the Kramers-Kronig relations, the real part of $P^{(0),r}$ is given by $\Re{P^{(0),r}}(\omega)={\cal P}\int d\omega' {\rho_P^{(0)}(\omega')}/(\omega-\omega')$, where $\rho_P^{(0)}=-\Im{P^{(0),r}}/\pi$ and ${\cal P}$ is the principal value. Therefore, the real part of ${P^{(0),r}}$ is negative for $\hbar \omega < \hbar \omega_p$, and positive for $\hbar \omega > \hbar \omega_p$ [red line in Fig.~\ref{Spectra}(a)].\par
Regarding $L^r$, the processes of the molecular absorption are classified into two kinds: the first one is accompanied by the vibrational transitions while the other is not (inelastic and elastic processes, respectively). From the spectral function of the molecule $\rho_L=-\Im{L^r}/\pi$, it is confirmed that for $\lambda^2<1$, the elastic process is dominant over the inelastic processes.~\cite{Mahan2000} Therefore the real part of ${L^r}$ is negative when the energy $\hbar \omega$ is lower than the energy for the elastic process $E_\mathrm{elastic}$ (1.87~eV), where the shift of the peak position from $\epsilon_{ex}$ is scaled as $V^2$.~\cite{Miwa2013} The real part, $\Re{L^r}$, tends to be positive in the energy range $ \hbar \omega > E_\mathrm{elastic}$ [red line in Fig.~\ref{Spectra}(b)].\par
Therefore, the product $\Re{P^{(0),r}} \Re{L^r}$ gives a positive contribution to the molecular luminescence intensity when $\hbar \omega$ is lower than both $E_\mathrm{elastic}$ and $\hbar \omega_p$. Thus, the interference between the processes of the molecular absorption and the re-absorption by the surface plasmons suppresses these energy absorption processes in this energy range. The product $\Re{P^{(0),r}} \Re{L^r}$ tends to be negative when $\hbar \omega$ is between $E_\mathrm{elastic}$ and $\hbar \omega_p$. Hence, the interference tends to enhance the energy absorption processes in this energy range. When the energy is higher than both $E_\mathrm{elastic}$ and $\hbar \omega_p$, the interference tends to suppress the energy absorption processes. This trend can be seen even when the energy of the surface plasmon mode $\hbar \omega_p$ is varied. Thus, the structure near 1.81~eV in $J_P$ (red line in Fig.~\ref{PL_Each}), which looks like a peak structure, can be attributed to the interference between these absorption processes.}}
\par
The peak and dip structures in $J_P$ can be seen in the recent experimental result.~\cite{Schneider2012} Moreover, our results show that the re-absorption by the surface plasmons leads to the dent structure in $J_P$. Thus, in addition to the molecular luminescence and absorption, the re-absorption by the surface plasmons plays an important role in determining the luminescence-spectral profiles of the surface plasmons. Also, it is found that the effects of the interference between the processes of the molecular absorption and the re-absorption by the surface plasmons can be seen in $J_P$. We expect that these results can be verified by experiments, for example by a comparison between the luminescence spectra acquired with a molecule-covers tip and with a clean metallic tip on a clean metal surface, which correspond to the red solid and green dashed lines in Fig. ~\ref{PL_Each}, respectively.
%====================================================================
%===== Molecular luminescence at high voltage =========
%====================================================================
\subsection{Molecular luminescence at the high bias voltage}
To analyze the luminescence-spectral profile of the molecule, contributions from each term in Eq. (\ref{eq:EachML}) are shown in Fig.~\ref{ML_Each}.
%Figure (Molecular Luminescence, Each terms) %%%%%%%%%%
\begin{figure}
\centering
\includegraphics[width = 8.6cm] {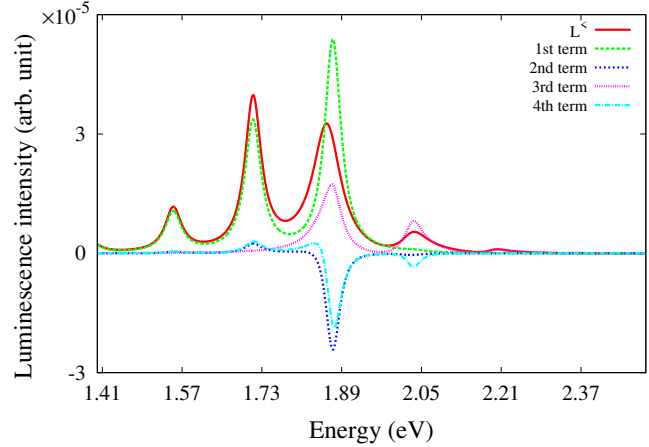}%
\caption{\label{ML_Each} (Color online) Calculation results for the imaginary part of each term on Eq. (\ref{eq:EachML}). Red solid line is the luminescence spectra of the molecule $J_L$. Green dashed, blue dotted, magenta dotted and cyan dashed-dotted lines are the product of $-1/\pi$ and the imaginary part of the first, second, third and fourth terms on the right side of Eq. (\ref{eq:EachML}), respectively. The bias voltage is $V_\mathrm{bias}=2.5~\mathrm{V}$}
\end{figure}
%%%%%%%%%%%%%%%%%%%%%%%%%%%%%%
% 1st is Radiative decay from ground vibrational level. 3rd is energy transfer from the plasmon to the molecule. 
The contribution from the first term on the right side of Eq.~(\ref{eq:EachML}) (green dashed line) leads to peak structures near 1.87~eV, 1.71~eV, 1.55~eV and so on. These structure correspond to the molecular luminescence associated with the transitions from the ground vibrational level for the excited electronic state to the vibrational levels for the ground electronic state of the molecule. The physical interpretation of the processes is that since the electronic and vibrational part in $L_b$ are decoupled, the creation (annihilation) of the molecular exciton induced by the absorption (emission) of the surface plasmons is not accompanied by the vibrational motions. Hence the first term $L_b^<$ expresses the molecular luminescence associated with the transitions from the ground vibrational level where the molecular vibration is initially distributed.\par
The third term (magenta dotted line) leads to peak structures near 2.05~eV, 2.21eV, and 2.37~eV. The position of these structures correspond to the energy of transitions from the excited vibrational levels for the excited electronic state to the vibrational levels for the ground electronic state of the molecule. The third term describes the processes, where the motion of the molecular exciton is accompanied by the vibrational motions. The molecule is excited to the vibrational levels for the excited electronic state by the absorption of the surface plasmons, and then de-excited to the ground vibrational level for the ground electronic state by subsequent emission of the surface plasmons. Thus, the third term leads to the luminescence, which are attributed to the radiative decay from the vibrational excited levels for the excited electronic state (hot luminescence).~\cite{Dong2010,Schneider2012}\par
The second and fourth terms give the processes, where the energy of the molecular exciton is absorbed by the surface plasmons and re-absorbed by the molecule. Effects of interference between the processes of the absorption by the surface plasmons and the re-absorption by the molecule can be seen, as described in the previous subsection.
%====================================================================
%===== Molecular luminescence at low voltage =========
%====================================================================
\subsection{Molecular luminescence at the low bias voltage}
Figure~\ref{ML_Bias} shows the luminescence spectra of the molecule $J_L$ at various bias voltages lower than $\epsilon_{ex}/e$.
%Figure ( Bias voltage dependence)%%%%%%%%%%
\begin{figure}
\centering
\includegraphics[width = 8.6cm] {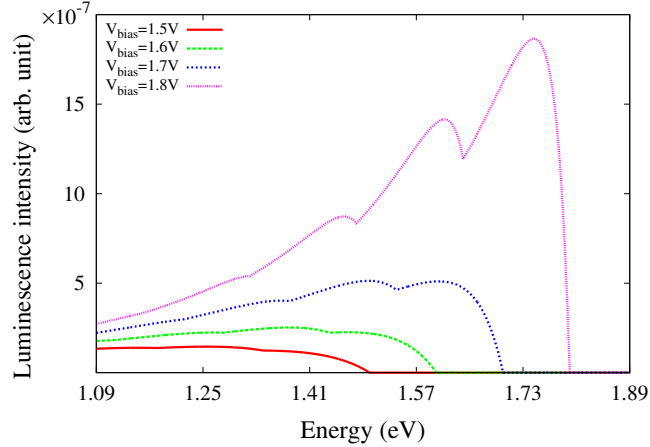}%
\caption{\label{ML_Bias} (Color online) Bias voltage $V_\mathrm{bias}$ dependence of the luminescence spectra of the molecule $J_L$ for $V_\mathrm{bias} < \epsilon_{ex}/e$. Red solid, green dashed, blue dotted and magenta dotted lines are $J_L$ at $V_\mathrm{bias}$=1.5~V, 1.6~V, 1.7~V and 1.8~V.}
\end{figure}
%%%%%%%%%%%%%%%%%%%%%%%%%%%%%%%%%%%%%%%%
Despite the fact that $eV_\mathrm{bias}$ is lower than the energy of the molecular exciton $\epsilon_{ex}$, the molecular luminescence is found to occur. The results indicate that the electron transitions in the molecule occur at this bias voltage. To analyze the mechanism for the occurrence of the electron transitions at $V_\mathrm{bias}<\epsilon_{ex}/e$, the spectral function of the molecule $\rho_L$ is calculated and shown in Fig.~\ref{Mol_Spe}.
%Figure (Spectral function of the molecule) %%%%%%%%%%%%%%%%%%%%
\begin{figure}
\centering
\includegraphics[ width = 8.6cm] {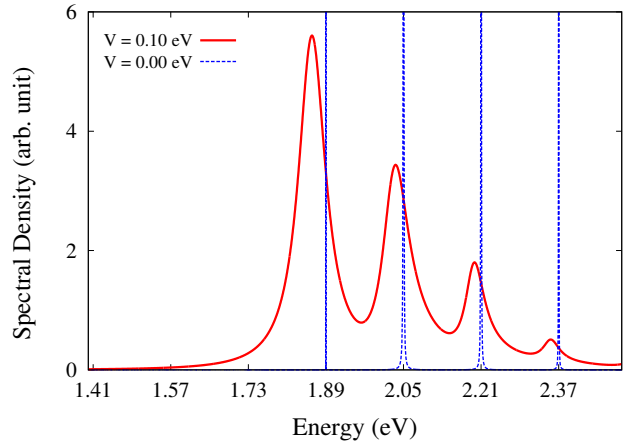}%
\caption{\label{Mol_Spe} (Color online) Spectral functions of the molecule for various values of $V$. Red solid and blue dashed lines are $\rho_L$ for $V=0.10~\mathrm{eV}$ and $V=0.00~\mathrm{eV}$, respectively. The bias voltage is $V_\mathrm{bias}=1.8~\mathrm{V}$}
\end{figure}
%%%%%%%%%%%%%%%%%%%%%%%%%%%%%%%%%%%%%%%%
Owing to the exciton-plasmon coupling $V$, the position and the width of the peaks in $\rho_L$ are shifted and broadened, respectively. The spectral intensities are found in the energy range lower than $\epsilon_{ex}$. It means that the excitation channels of the molecule arise in this energy range. Thus, the electron transitions in the molecule occur via the excitation channels resulting from the exciton-plasmon coupling, and give rise to the molecular luminescence at $V_\mathrm{bias} < \epsilon_{ex}/e$. Corresponding luminescence spectra can be seen in a recent experiment.~\cite{Fujiki2011a} The observed luminescence spectra shows peak structures near the energies of transitions in the molecule. The observed spectra also show the bias voltage dependence similar to Fig.~\ref{ML_Bias}. Thus, our results first reveal that the molecular luminescence at $V_\mathrm{bias} < \epsilon_{ex}/e$ are attributed to the electron transitions in the molecule via the excitation channels, that result from the exciton-plasmon coupling.
\par
Figure~\ref{ML_Upconv} shows the luminescence spectra of the molecule $J_L$ at $V_\mathrm{bias}=1.8~\mathrm{V}$.
%Figure ( Bias voltage dependence) %%%%%%%%%%%%%%%%%%%%
\begin{figure}
\centering
\includegraphics[ width = 8.6cm] {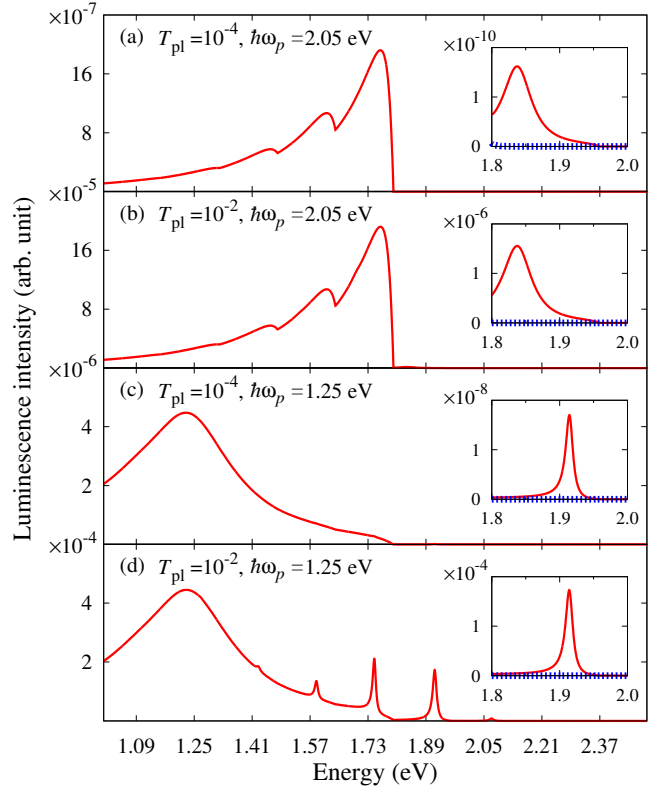}%
\caption{\label{ML_Upconv} (Color online) Luminescence spectra of the molecule $J_L$ at $V_\mathrm{bias} = 1.8~\mathrm{V}$. Insets: red solid and green dotted lines show $J_L$ for vibrational state in nonequilibrium and thermal equilibrium, respectively. (a) $T_\mathrm{pl}=10^{-4}$ and $\hbar\omega_p=2.05~\mathrm{eV}$, (b) $T_\mathrm{pl}=10^{-2}$ and $\hbar\omega_p=2.05~\mathrm{eV}$, (c) $T_\mathrm{pl}=10^{-4}$ and $\hbar\omega_p=1.25~\mathrm{eV}$, and (d) $T_\mathrm{pl}=10^{-2}$ and $\hbar\omega_p=1.25~\mathrm{eV}$.}
\end{figure}
%%%%%%%%%%%%%%%%%%%%%%%%%%%%%%%%%%%%%%%%
A peak structure is found in the energy range $\hbar \omega > eV_\mathrm{bias}$. Luminescence with the energy higher than $eV_\mathrm{bias}$ are called upconverted luminescence. One of the possible mechanisms for the occurrence of the upconverted luminescence is as follows: the electronic transitions of the molecule induced by the absorption and emission of the surface plasmons are accompanied by excitations of the molecular vibration, and the vibrational excitations assist the occurrence of the upconverted luminescence. The contribution of the vibrational excitations can be found in comparison with the vibrational state in thermal equilibrium, where the molecular vibration with the energy $\hbar \omega_0 = 0.16~\mathrm{eV}$ is distributed according to the Bose-Einstein distribution function $N_\mathrm{BE}$ at $T=80~\mathrm{K}$ and is therefore almost in the ground state.
\par
The dependence of luminescence spectra on $T_\mathrm{pl}$ and $\hbar \omega_p$ is also shown in Fig.~\ref{ML_Upconv}. The luminescence intensity in the energy range $\hbar \omega < eV_\mathrm{bias}$ is proportional to $T_\mathrm{pl}$, and the intensity of the upconverted luminescence is proportional to the square of $T_\mathrm{pl}$. As the energy of the surface plasmon mode $\hbar \omega_p$ is shifted to the low-energy side, the luminescence intensity increase. This is attributed to the fact that since the energy of the surface plasmon mode is lower than $eV_\mathrm{bias}$, the electron transitions in the molecule in the energy range $\hbar \omega < eV_\mathrm{bias}$ are enhanced by the surface plasmons.
\par
Figure~\ref{Occ} shows the $V_\mathrm{bias}$ dependence of the vibrational occupation number $n_b=\langle b^\dagger b \rangle_{\Tilde{H}}$ [red solid lines in Fig.~\ref{Occ}] and the population of the molecular exciton $n_e=\langle d^\dagger d \rangle_{\Tilde{H}}$ [blue dotted lines in Fig.~\ref{Occ}].
%Figure ( Bias voltage dependence) %%%%%%%%%%%%%%%%%%%%
\begin{figure}
\centering
\includegraphics[ width = 8.6cm] {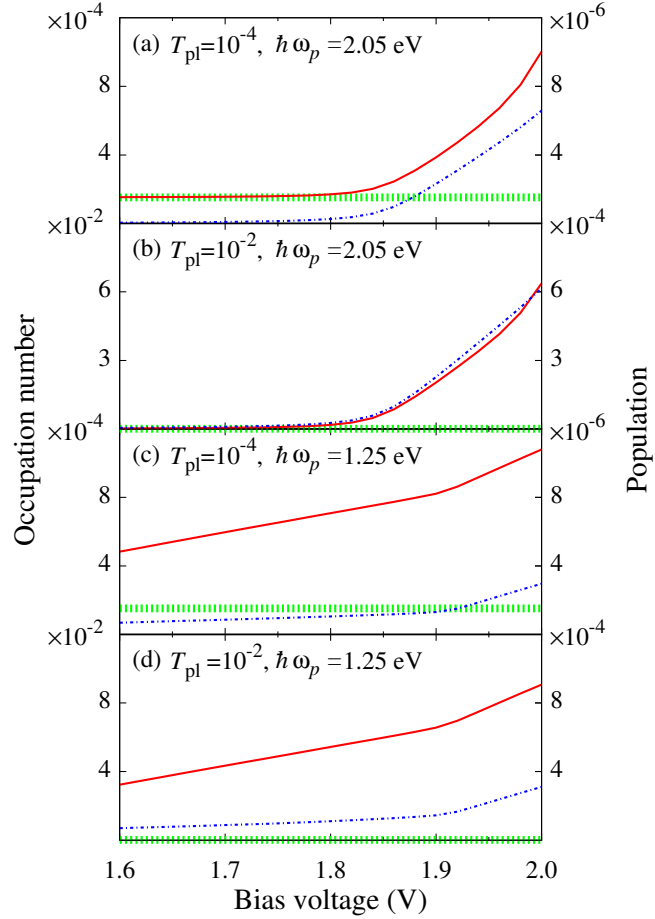}%
\caption{\label{Occ} (Color online) Bias voltage dependence of the vibrational occupation number $n_b$ and the population of the molecular exciton $n_e$. Red solid line indicates $n_b$. Green dashed line shows the vibrational occupational number for the vibrational state in thermal equilibrium $n_{b,\mathrm{TE}}$. Blue dotted line shows $n_e$. Here (a) $T_\mathrm{pl}=10^{-4}$ and $\hbar\omega_p=2.05~\mathrm{eV}$, (b) $T_\mathrm{pl}=10^{-2}$ and $\hbar\omega_p=2.05~\mathrm{eV}$, (c) $T_\mathrm{pl}=10^{-4}$ and $\hbar\omega_p=1.25~\mathrm{eV}$, and (d) $T_\mathrm{pl}=10^{-2}$ and $\hbar\omega_p=1.25~\mathrm{eV}$.}
\end{figure}
%%%%%%%%%%%%%%%%%%%%%%%%%%%%%%%%%%%%%%%%
It is confirmed that the vibrational excitations occur at $V_\mathrm{bias}=1.8~\mathrm{V}$.
Hence the vibrational excitations are found to assist the occurrence of the upconverted luminescence. 
The slope of $n_e$ changes at $V_\mathrm{bias}$ of approximately 1.85~{eV} for $\hbar \omega_p=2.05~\mathrm{eV}$ [Figs.~\ref{Occ}(a) and \ref{Occ}(b)] and at $V_\mathrm{bias}$ of approximately 1.90~{eV} for $\hbar \omega_p=1.25~\mathrm{eV}$ [Figs.~\ref{Occ}(c) and \ref{Occ}(d)]. At this bias voltage, the excitation channels of the molecule increase. The slope of the difference between $n_b$ and the vibrational occupational number for the vibrational state in thermal equilibrium $n_{b,\mathrm{TE}}$ changes at the same bias voltage. Therefore, the electron transitions in the molecule via the excitation channels resulting from the exciton-plasmon coupling give rise to the vibrational excitations at $V_\mathrm{bias} < \epsilon_{ex}/e$. It is found that $n_b$ is scaled by $\lambda^2 |V|^2 n_{ex} / \gamma$, where $\gamma$ represents the vibrational damping for $V=0~\mathrm{eV}$. Thus, even at $V_\mathrm{bias} < \epsilon_{ex}/e$, the vibrational excitations occur with the electron transitions in the molecule via the excitation channels resulting from the exciton-plasmon coupling and assist the occurrence of the upconverted luminescence.
%====================================================================
%===== Conclusion =========
%====================================================================
\section{Conclusion}
The luminescence properties of the molecule and the surface plasmons are greatly influenced by the interplay between the dynamics of the molecule and the surface plasmons.
The peak structures in $J_P$ arise due to the enhancement by the molecular modes and the dip structures due to the molecular absorption. The re-absorption by the surface plasmons leads to the dent structure in $J_P$. The interference between the processes of the molecular absorption and  the re-absorption by the surface plasmons is found to enhance or suppress the luminescence intensity of the surface plasmons, depending on the energy relative to $E_\mathrm{elastic}$ and $\hbar \omega_p$. The effects of the interference can be seen in $J_P$. 
Moreover, due to the exciton-plasmon coupling, the excitation channels of the molecule arise even in the energy range lower than $\epsilon_{ex}$. The electron transitions via these excitation channels give rise to the molecular luminescence at $V_\mathrm{bias} < \epsilon_{ex}/e$. It is also found that even at $V_\mathrm{bias} < \epsilon_{ex}/e$, the vibrational excitation occur with the electron transitions via these excitation channels and assist the occurrence of the upconverted luminescence.\par
Our results will pave the way for understanding the dynamics in the system where multiple bosonic particles (e.g. molecular exciton, vibron, surface plasmon, etc.) are coupled to each other. The analysis of the molecular and plasmon luminescence may contribute to the development of plasmon enhanced organic light-emitting diode (OLED) devise and high-resolution spectroscopy and sensing. Especially, the upconverted luminescence may open up new opportunities to increase the luminescence intensity in the high-energy region and explore highly efficient blue light-emitting materials for OLED devices.

%====================================================================
%===== Acknowledgment =========
%====================================================================
\begin{acknowledgment}
This work is supported in part
by MEXT
through the G-COE program,
Grant-in-Aid for Scientific Research on Innovative Areas Program
(2203-22104008) and Scientific Research (c) Program (22510107).
It was also supported in part by JST
through ALCA Program
and Strategic Japanese-Croatian Cooperative Program on Materials Science.
Some of the calculations presented here were performed using
the ISSP Super Computer Center, University of Tokyo.
The authors are deeply grateful to Professor Wilson Agerico Di\~{n}o
and Professor Hiroshi Nakanishi in Osaka University for useful discussions.
\end{acknowledgment}
%====================================================================
%===== Appendix =========
%====================================================================
%\appendix
%\section{}
%
%
%\bibliographystyle{jpsj}
%\bibliography{References}

\begin{thebibliography}{10}
\bibitem{foot2}
It should be noted that a more general model for the molecular vibrations,
  e.g., the one taking into account multiple vibrational modes and/or an
  intramolecular anharmonicity, can be found in the literature,~$^{41,42)}$
  %~\cite{Kasai1993,Kasai2001} and effects of the multiple vibrational modes
  and/or the intramolecular anharmonicity will be investigated in future works.
\bibitem{foot}
For example, when the molecule is directly adsorbed on the tip surface, the
  molecular exciton has a finite lifetime due to the energy transfer to the
  metal tip. This effect can be taken into account by introducing a relaxation
  constant $\Gamma_{ex}$. The lifetime of the molecular exciton for
  $V=0~\mathrm{eV}$ is proportional to $1/\Gamma_{ex}$. When $\Gamma_{ex}$ is
  large, the luminescence intensities of the molecule are expected to be much
  lower than the surface plasmons. In this case, the photon emission rate is
  approximately proportional to $P^<$. Although $\Gamma_{ex}$ has not been
  taken into account here, we have confirmed that $\Gamma_{ex}$ leads only to
  variation in the luminescence intensities and the peak width in $B_L$. Thus,
  the luminescence spectra detected experimentally are expected to correspond
  to $B_P$.
\end{thebibliography}

\end{document}